\documentclass[pra,superscriptaddress,noshowpacs,epsf]{revtex4}
\usepackage{graphicx}
\usepackage{amssymb}
\usepackage{comment}
\usepackage{amsmath}
\usepackage{color}

\begin{document}

\title{Adiabatic Mach-Zehnder interferometer in dipolar spin-1 condensate}
\author{Yixiao Huang}
\email{yxhuang1226@gmail.com}
\affiliation{School of Science, Zhejiang University of Science and Technology, Hangzhou,
Zhejiang, 310023, China}
\author{Wei Guo}
\affiliation{Zhejiang Institute of Modern Physics, Department of Physics, Zhejiang
University, Hangzhou 310027, China}
\author{Ning-Ju Hui}
\affiliation{Department of Applied Physics, Xi'an University of Technology, Xi'an 710054, China}
\author{Zheng-Da Hu}
\email{huyuanda1112@jiangnan.edu.cn}
\affiliation{School of Science, Jiangnan University, Wuxi 214122, China}
\affiliation{Jiangsu Provincial Research Center of Light Industrial Optoelectronic Engineering and Technology}

\begin{abstract}
Mach-Zehnder interferometer, a powerful tool for a wide variety of measurements,
has been realized with Bose-Einstein condensates in recent experiments.
In this report, we propose and analyze a realizable scheme for performing
a Heisenberg-limited Mach-Zehnder interferometry with dipolar spin-1 condensate.
Based upon adiabatic processes of sweeping the transverse magnetic field,
we demonstrate a perfect phase transition, which accomplishes the beam splitter, phase shifter and recombiner as for a Mach-Zehnder interferometer.
The attractive dipolar interaction ensures the existence of a path-entangled state
which enhances the phase measurement precision to the Heisenberg limit.
We also discuss the spin-$1$ squeezing induced in the adiabatic passage and
show that the squeezing parameter attains its minimal value near the point of saturation field.
%The proposed interferometry scheme can operate
%both for the order of ten and thousands atoms, which may be realizable in
%the experiment with current technology.
\end{abstract}

\maketitle

%\affiliation{Department of Physics and Center of Theoretical and Computational Physics,
%The University of Hong Kong, Hong Kong, China}

Quantum interference, one of the most fundamental and challenging principles in quantum mechanics, has attracted continuous attention in both theoretical and experimental studies~\cite{J. Rarity,Y. Ji,W. Oliver}.
It has been shown that quantum interference could be used to implement high-precision measurement, quantum information processing, etc.
A basic device capable of performing quantum interferometry is the Mach-Zehnder (MZ) interferometer,
which has two beam splitters for splitting the input states and recombining the output states, respectively.
A parameter $\phi $ that defined as the phase difference between the two beam splitters can be induced.
For the conventional Mach-Zehnder interferometry via unentangled states,
the quantum measurement can only reach the standard quantum limit or the shot noise limit, proportional to $1/\sqrt{N}$, where $N$ is the particle number~\cite{V. Giovannetti}. By contrast, the measurement precision can be enhanced to the
Heisenberg limit via entangled states~\cite{V. Giovannetti1,Yixiao Huang,J. Est,L. Pezz,Brukner,C. Invernizzi,J. Ma1,J. J. Bollinger,J. P. Dowling}.
An excellent candidate is the maximally entangled state $(\left\vert N\right\rangle _{a}\left\vert 0\right\rangle_{b}+\left\vert 0\right\rangle _{a}\left\vert N\right\rangle _{b})$/$\sqrt{2}$, which is called N-body GHZ state, or the so-called NOON state in quantum
optics with an equal-probability superposition of all $N$ particles in one of the two different paths.
The use of entangled ions for  high precision metrology have been studied both theoretically and experimentally~\cite{R. Huesmann,C. A. Sackett,D. Leibfried}.

On the other hand, the experimental realization of Bose-Einstein condensates (BECs) provides an ideal platform
to simulate quantum interferometry owing to their macroscopic coherence properties~\cite{M. R. Andrews,E. W. Hagley,I. I.
Bloch,A. Perrin,Yixiao Huang5}.
Since the first observation of BEC interference in the experiment~\cite{M. R. Andrews},
various BEC interferometers have been developed in the past few years.
Recently, BEC interferences were performed by using Bragg beams in ballistic expansion with atoms freely propagating in a guide~\cite{E. W. Hagley}.
%By using Bragg beams in ballistic expansion
%with atoms freely propagating in a guide, quantum interference with BECs
%were performed in the recent experiment~\cite{E. W. Hagley}.
Splitting a single trapped BEC into two separated clouds in a double well
potentials, the quantum coherence and interference were observed using
ballistic expansion after switching off the trapping potential~\cite{Y.
Shin,T. Schumm,G. B. Jo,G. B. Jo1,Baumga}.
In contrast to the schemes of photons and trapped ions which are subject to limited numbers
of ions or the requirement for individual addressing~\cite{R. Huesmann,C. A. Sackett,D. Leibfried}.
The possibility of performing a Heisenberg-limited MZ interferometry has been demonstrated with an ensemble of thousands of atoms in BECs~\cite{Pezze,J. Grond,J. Grond1,C. Lee,Berrada T}.

In this work, we propose and analyze a realizable scheme for performing a Heisenberg-limited Mach-Zehnder interferometry
in spin-$1$ condensate with dipolar interaction.
The system of the dipolar spin-1 condensate is described by a uniaxial magnetic model with ferromagnetic (FM) interaction and magnetic field,
which has been used to simulate quantum magnetism and quantum phase transition via modifying the trapping geometry~\cite{Yixiao
Huang1,S. Yi,S. Yi2,Yixiao Huang3}. In our scheme, the system starts from a paramagnetic state dominated by a transverse magnetic field $H_{x}$.
We first adiabatically decrease $H_{x}$ to realize the beam splitter and phase shifter.
Then, the beam recombination is achieved by increasing $H_{x}$ from $0$ to a large value adiabatically,
which is the inverse process of the splitting. The adiabatic processes in our scheme perfectly connect the limit of transverse field $H_{x}$. Therefore, a pure GHZ state can be prepared adiabatically, which enhances the phase measurement precision to the Heisenberg limit.
Meanwhile, the phase shift can be extracted by measuring the population in the ground state and excited state by adiabatically increasing the transverse field $H_{x}$ to a large value in the recombining process.
We also discuss spin squeezing of the state generated in the adiabatic process and
show that the squeezing parameter attains its minimal value near the point of saturation field.
%With the help
%of off-resonant optical dipole traps, our proposed
%interferometry scheme can operate either for the order of $10$ or large
%particle numbers ($\sim 10^{4}$).

\textbf{Results }

\textbf{Model and ground state property.} We consider a spin $f=1$ condensate with $N$ atoms trapped in an axially symmetric potential.
For simplicity, we choose the symmetry axis to be the quantization axis $z$.
Under a uniform magnetic field $\mathbf{B}$, the second quantized Hamiltonian of the system without long-range magnetic dipolar interaction reads%
\begin{align}
\mathcal{H}_{0}=& \int d\mathbf{r}\hat{\psi}_{\alpha }^{\dagger }(\mathbf{r})%
{\bigg [}(-\frac{\hbar ^{2}\nabla ^{2}}{2M}+V_{\text{ext}}(\mathbf{r}))\hat{%
\psi}_{\alpha }(\mathbf{r})-g_{F}\mu _{B}\mathbf{B\cdot F}_{\alpha
\beta }\hat{\psi}_{\beta }(\mathbf{r}){\bigg ]}+\frac{c_{0}}{2}\int d\mathbf{%
r}\hat{\psi}_{\alpha }^{\dagger }(\mathbf{r})\hat{\psi}_{\beta }^{\dagger }(%
\mathbf{r})\hat{\psi}_{\alpha }(\mathbf{r})\hat{\psi}_{\beta }(\mathbf{r})
\notag \\
& +\frac{c_{2}}{2}\int d\mathbf{r}\hat{\psi}_{\alpha }^{\dagger }(\mathbf{r})%
\hat{\psi}_{\alpha ^{\prime }}^{\dagger }(\mathbf{r})\mathbf{F}_{\alpha
\beta }\cdot \mathbf{F}_{\alpha ^{\prime }\beta ^{\prime }}\hat{\psi}_{\beta
}(\mathbf{r})\hat{\psi}_{\beta ^{\prime }}(\mathbf{r})\text{,}  \label{Hsp}
\end{align}%
where $\hat{\psi}_{\alpha }(\mathbf{r})$ is the atomic field annihilation operator associated with atoms in the state $\left\vert f=1,m_{f}=\alpha
\right\rangle (\alpha =0,\pm 1)$, $\mathbf{F}$ denotes the angular momentum
operator, $\mu _{B}$ represents the Bohr magneton, and $g_{F}$ is the Land\'{e} $g$%
-factor. The mass of the atom is given by $M$ and the trapping potential $V_{%
\text{ext}}(\mathbf{r})$ is assumed to be spin independent. The collisional
interaction parameters are
\begin{align}
c_{0}& =4\pi \hbar ^{2}(a_{0}+2a_{2})/(3M), \nonumber\\
c_{2}& =4\pi \hbar ^{2}(a_{2}-a_{0})/(3M),
\end{align}%
where $a_{f}$ $(f=0,2)$ is the $s$-wave scattering length for spin-$1$ atoms in the combined symmetric channel of total spin $f$.

\begin{figure}[tbp]
\includegraphics[width=9cm,clip]{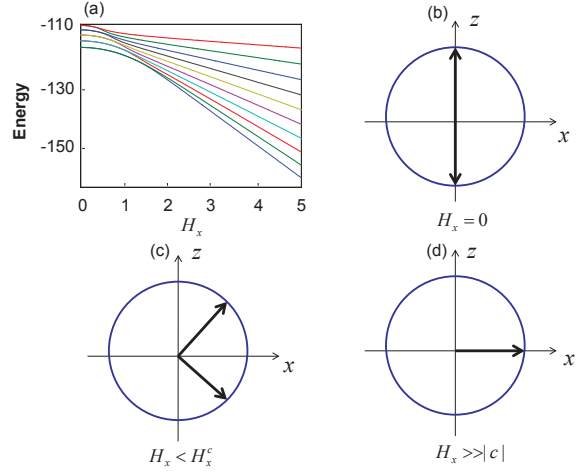} .
\caption{\textbf{Energy spectra and classical picture of the spin orientation.} (a) The energy spectra as a function of $H_{x}$ with $H_{z}=0$, $c=-0.1$, and $N=10$. (b)-(d) Classical picture of the spin orientation for different transverse fields with $H_{z}=0$, where the black arrowheads denote the spin orientation.}
\label{energy_N012}
\end{figure}

The Hamiltonian for the dipolar interactions reads%
\begin{align}
\mathcal{H}_{dd}=& \frac{c_{d}}{2}\int d\mathbf{r}\int d\mathbf{r}^{\prime }%
\frac{1}{\left\vert \mathbf{r}-\mathbf{r}^{\prime }\right\vert ^{3}}\times {%
\bigg [}\hat{\psi}_{\alpha }^{\dagger }(\mathbf{r})\hat{\psi}_{\alpha
^{\prime }}^{\dagger }(\mathbf{r}^{\prime })\mathbf{F}_{\alpha \beta }\cdot
\mathbf{F}_{\alpha ^{\prime }\beta ^{\prime }}\hat{\psi}_{\beta }(\mathbf{r})%
\hat{\psi}_{\beta ^{\prime }}(\mathbf{r}^{\prime })-3\hat{\psi}_{\alpha
}^{\dagger }(\mathbf{r})  \notag \\
& \times \hat{\psi}_{\alpha ^{\prime }}^{\dagger }(\mathbf{r}^{\prime })(%
\mathbf{F}_{\alpha \beta }\cdot \mathbf{e)(F}_{\alpha ^{\prime }\beta
^{\prime }}\cdot \mathbf{e)}\hat{\psi}_{\beta }(\mathbf{r})\hat{\psi}_{\beta
^{\prime }}(\mathbf{r}^{\prime }){\bigg ]}\text{,}  \label{Hdd}
\end{align}%
where $\mathbf{e}=(\mathbf{r}-\mathbf{r}^{\prime })/\left\vert \mathbf{r}-\mathbf{r}^{\prime }\right\vert \mathbf{\ }$ is a unit vector,
and $c_{d}=\mu_{0}g_{F}^{2}\mu _{B}^{2}/4\pi $ is the dipolar interaction parameter with $\mu _{0}$ being the vacuum magnetic permeability.
The total Hamiltonian is then given by $\mathcal{H}=\mathcal{H}_{0}+\mathcal{H}_{dd}$.
For both the $^{87}$Rb and $^{23}$Na atoms, we have $|c_{2}|\ll c_{0}$ and $c_{d}\lesssim 0.1|c_{2}|$.
Under these conditions, the single mode approximation, which describes the atoms in different spin states with the same wave function $\phi(\mathbf{r})$,
is expected to be valid, and then the field operator can be decomposed as~\cite{S. Yi2,C. K. Law,S. Yi3}%
\begin{equation}
\hat{\psi}_{\alpha }(\mathbf{r})\simeq \phi (\mathbf{r})\hat{a}_{\alpha }%
\text{,}  \label{SMA}
\end{equation}%
where $\hat{a}_{\alpha }$ is the annihilation operator of spin component $\alpha $.
Under the single mode approximation, the Hamiltonian of the system (with constant terms dropped) can be remarkably reduced to~\cite{S. Yi2}%
\begin{equation}
\mathcal{H=(}c_{2}^{\prime }-c_{d}^{\prime })\mathbf{\hat{L}}%
^{2}+3c_{d}^{\prime }(\hat{L}_{z}^{2}+\hat{n}_{0})-g_{F}\mu _{B}\mathbf{B}%
\cdot \mathbf{\hat{L}}\text{,}
\end{equation}%
where $\mathbf{\hat{L}=}\hat{a}_{\alpha }^{\dagger }\mathbf{F}_{\alpha \beta
}\hat{a}_{\beta }$ is the total angular momentum operator, $\hat{L}_{z}$ is
its $z$-component, and $\hat{n}_{0}=\hat{a}_{0}^{\dagger }\hat{a}_{0}$.
The parameters $c_{2}^{\prime }$ and $c_{d}^{\prime }$ are the rescaled collisional and
dipolar interaction strengths, respectively, which are given by
\begin{eqnarray}
c_{2}^{\prime } &=&\frac{c_{2}}{2}\int d\mathbf{r}|\phi (\mathbf{r})|^{4}%
\text{,} \nonumber\\
c_{d}^{\prime } &=&\frac{c_{d}}{4}\int d\mathbf{r}d\mathbf{r}^{\prime }\frac{%
|\phi (\mathbf{r})|^{2}|\phi (\mathbf{r}^{\prime })|^{2}}{\left\vert \mathbf{%
r}-\mathbf{r}^{\prime }\right\vert ^{3}}\left( 1-3\cos ^{2}\theta_{e}\right),
\end{eqnarray}%
with $\theta _{e}$ being the polar and azimuthal angles of ($\mathbf{r}-\mathbf{r}^{\prime }$). The sign of $c_{2}^{\prime }$ is determined by the
type of atoms: $^{87}$Rb ($c_{2}^{\prime }<0$) and $^{23}$Na ($c_{2}^{\prime}>0$).
The sign and the magnitude of dipolar interaction strength $c_{d}^{\prime }$ can be tuned via modifying the trapping geometry (see Methods).
Considering the case of $c_{2}<0$ for $^{87}$Rb atoms and rescaling the Hamiltonian by using $|c_{2}^{\prime }|$ as the energy unit,
the dimensionless Hamiltonian reduces to~\cite{S. Yi2}%
\begin{equation}
\mathcal{H=(}-1-c)\mathbf{\hat{L}}^{2}+3c(\hat{L}_{z}^{2}+\hat{n}_{0})-%
\mathbf{H}\cdot \mathbf{\hat{L}}\text{,}  \label{Hami}
\end{equation}%
where $c=c_{d}^{\prime }/|c_{2}^{\prime }|$ and $\mathbf{H}=g_{F}\mu _{B}%
\mathbf{B}$.

The ground state of the system sensitively depends on the parameters $c$ and $\mathbf{H}$.
For the transverse magnetic field $H_{x}/|c|\gg 1$, the ground state $\left\vert \Psi _{G}\right\rangle $ is a spin coherent state $\left\vert N,N\right\rangle _{x}$.
In the limit of weak transverse magnetic field $H_{x}/|c|\ll 1$, the ground state relies on $c$.
If $c>0$ and $H_{x}\rightarrow 0$, the ground state approaches to $\left\vert N,0\right\rangle $ for even $N$.
If $c<0$, the ground state $\left\vert \Psi _{G}\right\rangle $ and the first excited state $\left\vert \Psi _{E}\right\rangle $
become degenerate in the limit of $H_{x}\rightarrow 0$.
In such a limit, the ground state and the first excited state are the lowest spin state $\left\vert N,N\right\rangle $
and the highest spin state $\left\vert N,-N\right\rangle $, respectively.
Beside the magnetic field along the $x$-axis, if we also consider a non-zero longitudinal magnetic field (i.e., $H_{z}\neq 0$),
the degeneracy between $\left\vert \Psi _{G}\right\rangle $ and $\left\vert \Psi_{E}\right\rangle $ will be destroyed.
In Fig.~\ref{energy_N012}, we plot the energy spectra and the classical picture of the spin orientation for different magnetic fields.
As shown in Fig.~\ref{energy_N012} (a), when the transverse field $H_{x}$ reduces to zero,
the energy of the ground and first excited states become degenerate.
In the limit of\ large particle numbers $N\gg 1$, the energy of the Hamiltonian (\ref{Hami})
becomes $E(\vartheta )=3cN^{2}\cos ^{2}\vartheta -HN\sin \vartheta $,
where $\vartheta $ is the polar angle of $\mathbf{L}$.
Minimizing the energy with respect to $\vartheta$ yields the optimal value for $\vartheta =\pi /2$ when $H_{x}\gg |c|$,
and $\vartheta =0$ or $\pi $ when $H_{x}=0$.
The critical field strength is given by $H_{x}^{c}=-6Nc $~\cite{S. Yi}.
As shown in Fig.~\ref{energy_N012} (b)-(d), when $H<H_{x}^{c}$,
the ground state is doubly degenerate, while if $H>H_{x}^{c}$, the two degenerate states collapse into one and the system is fully polarized by
the transverse field. %As shown in Fig. \ref{energy_N012},
%the classical picture of the spin orientation is plotted for different transverse field.

\begin{figure}[tbp]
\includegraphics[width=8.5cm,clip]{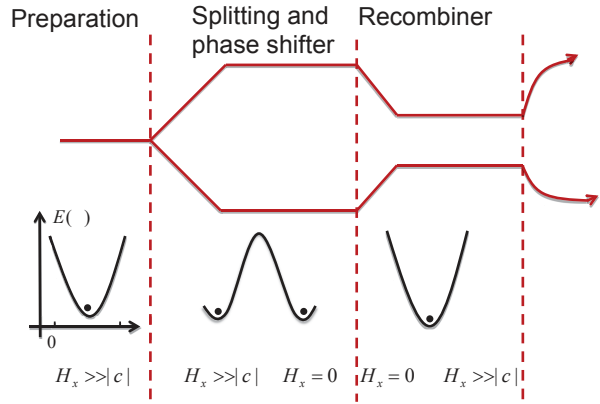}
\caption{\textbf{Mach-Zehnder interferometer of BEC-type.} Schematic diagram of the
Mach-Zehnder interferometer (top) and energy curves for different transverse field limits (bottom).}
\label{scheme}
\end{figure}

Now we propose a full Mach-Zehnder interferometer in the dipolar spin-$1$ condensate with $c<0$.
Our interferometric scheme relies on the coherent splitting and recombination of the BEC in the tunable magnetic field.
In the top of Fig.~\ref{scheme}, we show the schematic of Mach-Zehnder interferometer.
The system is initialized with strong transverse field. Then the ground state and first excited state can be considered as two paths of an interferometer, which accomplishes the beam splitting. With a weak longitudinal magnetic field, a phase shift between the two arms is imprinted by
slowly decreasing $H_{x}$ from $H_{x}\gg |c|$ to zero.
Finally, the recombination is achieved by conversely increasing $H_{x}$ from zero to $H_{x}\gg |c|$.
The energy curves for the three different transverse field limits, shown in the bottom of Fig.~\ref{scheme}, correspond to processes of the preparation, splitting, and recombining, respectively. In what follows, we present the details of the scheme.

\begin{figure}[tbp]
\includegraphics[width=8.8cm,clip]{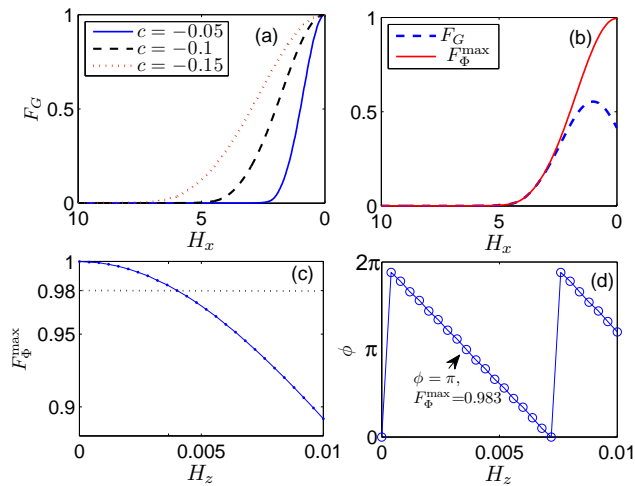}
\caption{\textbf{Effects of external fields.} The transverse field dependence of the fidelity for (a) $H_{z}=0$ and (b) $H_{z}=0.002$.
(c) The maximum fidelity $F_{\Phi }^{\max }$ and (d) the phase shift $\phi $ as functions of $H_{z}$ in the limit of $H_{x}\rightarrow 0$.
The parameters are chosen as $N=10$, $c=-0.1$ and the linear field sweeping rate $\mathrm{d}H_{x}/\mathrm{d}t=0.08$.}
\label{overlap_GHZ}
\end{figure}

\textbf{Splitter and phase shifter.} Initially, the ground state is prepared in the strong transverse magnetic field ($H_{x}^{(i)}\gg |c|$) and
then the transverse magnetic field is linearly swept at a constant rate, i.e., $H_{x}(t)=H_{x}^{(i)}-vt$ for $t\geqslant 0$.
In Fig.~\ref{overlap_GHZ} (a), we plot the overlap (the so-called fidelity $F_{G}=|\left\langle \Psi (t)\right. \left\vert \text{GHZ}\right\rangle |^{2}$) between the maximally spin entangled GHZ state and the generated state as a function of the time dependent magnetic field $H_{x}$.
It is clearly observed that the fidelity $F_{G}$ between the generated and GHZ states reaches its maximum ($F_{G}=1$) as $H_{x}\rightarrow 0$.
This implies that a path-entangled state $(\left\vert N,N\right\rangle +\left\vert N,-N\right\rangle )/\sqrt{2}$ can be created in the limit of $H_{x}\rightarrow 0$. Thus, the beam splitter in the first process also provides a route to produce a kind of macroscopic entangled state.

In the process above, if we additionally apply a weak longitudinal magnetic field $H_{z}$, Landau-Zener tunneling will be induced.
In a recent experiment of superconducting flux qubit, the Mach-Zehnder interferometry has been demonstrated by utilizing Landau-Zener tunneling as a beam splitter~\cite{W. Oliver}.
In the present scheme, one can instead obtain a path-entangled state $\left\vert \Phi (\phi )\right\rangle_{\mathrm{GHZ}} =(\left\vert
N,N\right\rangle +e^{\mathrm{i}\phi }\left\vert N,-N\right\rangle )/\sqrt{2}$ with certain phase shift.
This can be concluded from Fig.~\ref{overlap_GHZ}(b), where the fidelity $F_{G }$ and the maximal fidelity $F_{\Phi }^{\max }=\max(|\left\langle \Psi (t)\right. \left\vert \Phi (\phi )\right\rangle_{\mathrm{GHZ}} |^{2})$ are plotted as functions of $H_{x}$ with $\phi$ ranging from $0$ to $\pi$.
It is displayed that the fidelity $F_{G }<1$ while the maximal fidelity $F_{\Phi }^{\max }\simeq 1$ in the limit of $H_{x}\rightarrow 0$,
which indicates that the presence of the longitudinal magnetic field imprints a phase shift
between the spin states $\left\vert N,N\right\rangle$ and $\left\vert N,-N\right\rangle $.
%Here we note that the phase difference $\phi $ mainly comes from $H_{z}$ and the critical value $H_{x}^{c}$ depends on the parameters and the desired fidelity.
It is worth noting that the strength of the longitudinal magnetic field $H_{z}$ should be weak so as to achieve the desired high fidelity.
To illustrate this fact, the maximal fidelity $F_{\Phi }^{\max}=\max(|\left\langle \Psi (H_{x}=0,H_{z})\right. \left\vert \Phi (\phi
)\right\rangle_{\mathrm{GHZ}} |^{2})$ is plotted as a function of $H_{z}$ in Fig.~\ref{overlap_GHZ}(c). It is demonstrated that one can obtain the path-entangled state $(\left\vert N,N\right\rangle +e^{i\phi }\left\vert N,-N\right\rangle )/\sqrt{2}$ with the fidelity more than 0.98 by controlling the longitudinal magnetic field $H_{z}$ to be weak enough.
Moreover, to explore the exact dependence of the phase shift $\phi$ on the weak longitudinal magnetic field $H_{z}$, in Fig.~\ref{overlap_GHZ} (d),
we plot the phase shift $\phi$ as a function of magnetic field $H_{z}$ in the limit of $H_{x}\rightarrow 0$.
It can be seen that the phase shift $\phi$ exhibits a linear relation to the weak longitudinal magnetic filed $H_{z}$.
Based on such a result, the information of the phase shift may be extracted if we have the knowledge of $H_{z}$, and vice versa.
%In addition, controlling the magnetic field $H_{z}$ one
%can get the path-entangled states $(\left\vert N,N\right\rangle +e^{i\phi
%}\left\vert N,-N\right\rangle )/\sqrt{2}$ with high fidelity larger than
%0.98, see.

\textbf{Recombiner.} To extract the phase shift $\phi$ between the two different spin states,
one has to transform the phase information into the amplitude information of specific observable for the final state.
For our scheme, we may accomplish this procedure in the recombination process and monitor the populations in the two output paths.
The beam recombination can be achieved by a Landau-Zener tunneling as to adiabatically increase $H_{x}$ from $H_{x}\rightarrow 0$ to $H_{x}\gg |c|$,
which is the inverse process of splitting.
Finally, with the strong transverse magnetic field $H_{x}$, one can measure the populations of spin components $0$ and $\pm 1$ in the ground and the first excited states, which will show interference behaviors and relate to the phase shift $\phi$. For example, by inversely increasing the transverse field $H_{x}$ from $0$ to $10$ and removing the longitudinal magnetic field $H_{z}$,
we examine the fidelities of the final state $\left\vert \Psi (H_{x}=10)\right\rangle$ to the ground state $\left\vert \Psi_G\right\rangle$ and to the first excited state $\left\vert \Psi_E\right\rangle$ after the beam recombination process.
In ideal case, we find that $F_{0}=|\left\langle \Psi(H_{x}=10)\right. \left\vert
\Psi_G\right\rangle |^{2}=\cos ^{2}(\phi /2)$ and $F_{1}=|\left\langle \Psi (H_{x}=10)\right. \left\vert \Psi_E\right\rangle |^{2}=\sin ^{2}(\phi /2)$, respectively. It can be concluded that all of the particles will occupy the ground state if $\phi =2n\pi $ or will stay in the first excited state if $\phi =(2n+1)\pi$ with $n$ an integer.

\begin{figure}[tbp]
\includegraphics[width=8.8cm,clip]{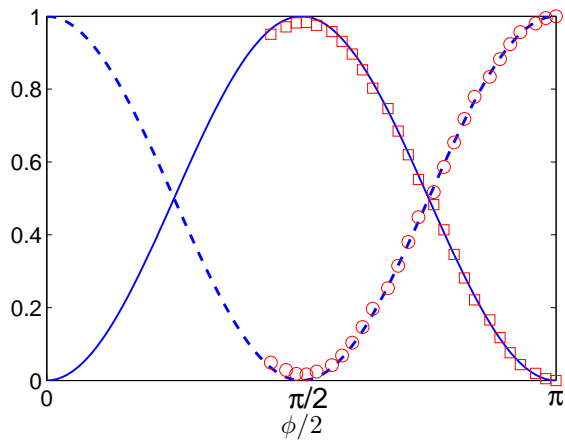}
\caption{\textbf{Fidelities of the output state.} Behaviors of Mach-Zehnder
interference in the output states with $N=10$, $c=-0.1$, and $H_{z}=0$. The
blue dashed and solid lines denotes the fidelity $F^{'}_{0}$ and $F^{'}_{1}$ with
initial state $\left\vert \Psi (\protect\phi )\right\rangle _{\mathrm{GHZ}}$, respectively.
The red cirles and squares correspond to $F_{0}$ and $F_{1}$ with initial
state generated in the process of splitting and phase shift, respectively.}
\label{Fidelity_recombiner_2}
\end{figure}

In Fig.~\ref{Fidelity_recombiner_2}, we plot the fidelities $F_{0}$ and $F_{1}$ as a function of $\phi $,
which shows a perfect behavior of Mach-Zehnder interference.
Here, we note that the state generated in the process of first beam splitting may not exactly be the phase-shifted GHZ state
$\left\vert \Phi (\phi )\right\rangle_{\mathrm{GHZ}}=(\left\vert
N,N\right\rangle +e^{i\phi }\left\vert N,-N\right\rangle )/\sqrt{2}$  (see Fig.~\ref{overlap_GHZ}(c)).
For comparison, we also demonstrate the result of the fidelities $F^{'}_{0}$ and $F^{'}_{1}$ by assuming that the
initial state for the recombing process is $\left\vert \Phi (\phi )\right\rangle_{\mathrm{GHZ}}$.
%generated in the process of first beam splitting.\textbf{\ }%
The numerical results states that when $\phi >\pi $, in which region the
maximum fidelity $F_{\Phi }^{\max }>0.983$ (see Fig.~\ref{overlap_GHZ}(d)), the fidelities $F^{'}_{0}$ and $F^{'}_{1}$ for the initial state phase-shifted GHZ state $\left\vert \Phi (\phi )\right\rangle_{\mathrm{GHZ}}$ agree well with that for the
initial state generated in the process of beam splitting and phase shift.
Thus, the phase shift can be extracted by measuring the populations for the ground state $\left\vert \Psi_G\right\rangle$ and  the first excited state $\left\vert \Psi_E\right\rangle$ after the beam recombination process.

The scheme shall be experimentally realizable with current technologies due to its advantages.
First, the system always remains in the minimal energy state throughout the whole process,
making it immune to the spontaneous emission induced decoherence existing in the electronically excited states.
Moreover, this scheme does not rely on a precise knowledge of system parameters such as particle numbers and interaction strengths.
%In our system, the particle number can be controlled
%from the order of $10$ particles or large particle numbers ($\sim 10^{4}$)
%with the off-resonant optical dipole traps~\cite{Stamper-Kurn}.
%\begin{equation*}
%\end{equation*}

%\begin{figure}[tbp]
%\includegraphics[width=8.5cm,clip]{fig5.eps}
%\caption{\textbf{Quantum Fisher information.} Quantum Fisher information as
%a function of $H_{x}$ with $N=20$, $c=-0.1$, and the linear field sweeping
%rat $dH_{x}/dt=0.08$.}
%\label{Fisher_information}
%\end{figure}

\begin{figure}[tbp]
\includegraphics[width=8.5cm,clip]{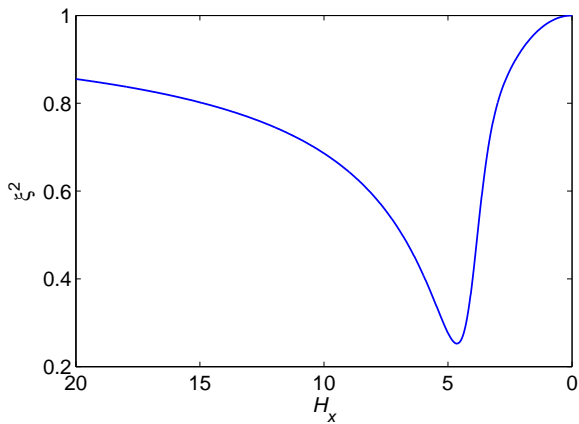}
\caption{\textbf{Spin squeezing.} Spin squeezing parameter $\protect\xi^{2}$
as a function of $H_{x}$ with $N=10$, $c=-0.1$, and the linear field
sweeping rate $\mathrm{d}H_{x}/\mathrm{d}t=0.08$.}
\label{Spin squeezing}
\end{figure}

\textbf{Spin squeezing.} Finally, we would like to discuss the generation of spin-$1$ squeezing with the scheme above.
Spin squeezing, which has potential applications in quantum physics such as atom interferometers, high precision atom clocks and quantum information,
is quantified by the following parameter~\cite{M. Kitagawa,J. Ma}
\begin{equation}
\xi ^{2}=\frac{4\min (\Delta S_{n_{\perp }})^{2}}{N}
\end{equation}%
with $(\Delta S_{n_{\perp }})^{2}=\left\langle S_{n_{\perp}}^{2}\right\rangle -\left\langle S_{n_{\perp }}\right\rangle ^{2}$.
The subscript $n_{\perp }$ refers to an arbitrary axis perpendicular to the mean spin $S$.
The inequality $\xi ^{2}<1$ indicates the state is spin squeezed.
For the GHZ state, since $\left\langle S_{x,y,z}\right\rangle=0 $,
which means that the mean spin vanishes and then the spin squeezing parameter $\xi ^{2}$ has no specific definition.
However, the spin squeezing can be induced in the process of beam splitting and phase shift.
In addition, a phase transition will occur as the transverse magnetic field sweeps from $H_{x}\gg |c|$ to $H_{x}\rightarrow 0$, and the
corresponding spin squeezing for the ground state exhibits a sharp transition phenomenon.
In Fig.~\ref{Spin squeezing}, the spin squeezing $\xi ^{2}$ is plotted as a function of time dependent $H_{x}$.
It is shown that $\xi ^{2}$ attains a minimal value near the position $H_{x}\approx H_{x}^{c}$, indicating a maximal spin squeezing.
The above result indicates that the splitting process can also generate spin squeezing.
The fact that critical points are correlated with the extremum values of spin squeezing is not a coincidence.
Quantum entanglement and quantum correlation may also reach their extremum values near (or at) the critical points of quantum phase transitions
in many spin systems~\cite{X. Wang,T. J. Osborne,J. Vidal}.
However, a full understanding of spin-$1$ squeezing in an adiabatic process is still lack.
With the tunability of dipolar interaction strength and magnetic field,
the dipolar spinor condensate appears to be an ideal system to generate spin-$1$ squeezing.%

\begin{figure}[tbp]
\includegraphics[width=8.5cm,clip]{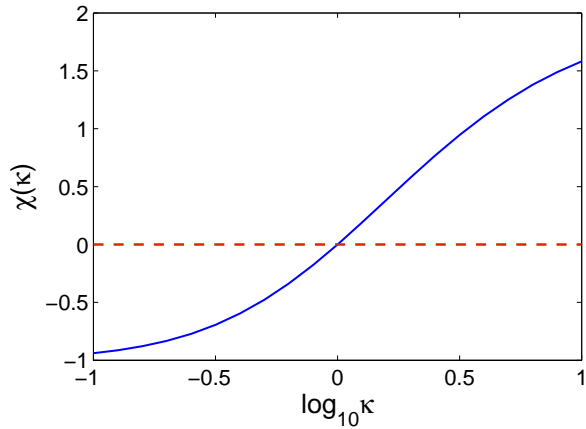}
\caption{\textbf{Dipolar interaction.}
The parameter $\chi (\kappa )$ of dipolar interaction  as a function of the trapping geometry parameter $\kappa$.}
\label{dipole_parameter}
\end{figure}

\textbf{Discussion}

In conclusion, we have presented a simple and realizable scheme for performing Heisenberg-limited MZ interferometry
in the dipolar spin-$1$ condensate.
The beam splitter, phase shifter and beam recombiner are achieved by adiabatically sweeping the transverse magnetic field.
In the beam splitting process, the transverse field adiabatically reduces to zero and the system generate a GHZ state.
A phase shift can be generated with a nonzero longitudinal field $H_{z}$.
In the recombining process, the phase shift can be extracted by measuring the population in the ground state and the first excited state with the
transverse field $H_{x}$ adiabatically increasing to $H_{x}\gg |c|$.
We also investigate the generation of spin-$1$ squeezing in the process of beam splitting
and phase shift and show that the squeezing attains a minimum value near the point of saturation field.
%The scheme of the MZ interferometry we
%present can operate both for the order of $10$ particles or large particle
%numbers ($\sim 10^{4}$) with the current experiment technology.

\textbf{Methods}

To calculate the rescaled collisional and dipolar interaction strengths $c_{0,2}^{\prime }$ and $c_{d}^{\prime }$,
we assume that the condensate wave function has a Gaussian form $\phi (r)=\pi ^{-3/4}\kappa^{1/4}\mathrm{e}^{-(x^{2}+y^{2}+\kappa ^{2}z^{2})/2}$,
and then obtain%
\begin{align}
c_{0,2}^{\prime } =&\frac{c_{0,2}}{2(2\pi )^{3/2}q_{r}^{2}q_{z}},\nonumber \\
c_{d}^{\prime } =&\frac{c_{d}}{6\sqrt{2\pi }q_{r}^{2}q_{z}}\frac{2\kappa^{2}+1-3\kappa ^{2}H(\kappa )}{\kappa ^{2}-1}
\end{align}%
with $\kappa =q_{r}/q_{z}$ and $H(\kappa )=\tanh^{-1}\sqrt{1-\kappa ^{2}}/\sqrt{1-\kappa ^{2}}$~\cite{S. Yi,S. Yi2}.
Therefore, we have $c\equiv c_{d}^{\prime }/c_{2}^{\prime }=2\pi c_{d}\chi (\kappa )/(3c_{2})$
with $\chi (\kappa )=[2\kappa ^{2}+1-3\kappa ^{2}H(\kappa )]/(\kappa^{2}-1)$.
In Fig.~\ref{dipole_parameter}, we plot the parameter $\chi(\kappa )$ of dipolar interaction as a function of the trapping geometry $\kappa $.
It can be seen that the value of $\chi(\kappa)$ can be tuned from $-1$ to $2$ by changing the trapping geometry parameter $\kappa $ from $0.1$ to $10$.
When $\kappa < 1$, the dipolar interaction is attractive, and if $\kappa > 1$,
the dipolar interaction is repulsive. The dipolar interaction disappears when $\kappa=1$.

\vspace{.5cm} \noindent {\large \textbf{Acknowledgments}}

\vspace{.01cm} \noindent  Y.H. acknowledges the Fundamental Research Funds for and the Central Universities
(Grant No. F701108F01). N.-J.H. acknowledges the National Natural Science Foundation of Special Theoretical Physics (Grant No.~11447217). Z.-D.H. acknowledges the natural science foundation of Jiangsu province of China (Grant No. BK20140128), the National Natural Science Foundation of Special Theoretical Physics (Grant No.~11447174), the National Natural Science Foundation of China (Grant No.~11504140) and the Fundamental Research Funds for the Central Universities (Grant No.~JUSRP51517).

\vspace{.5cm} \noindent {\large \textbf{Author contributions}}

\vspace{.01cm} \noindent Y.H. conceived the research and wrote the main manuscript text.  W.G., N.-J.H. and Z.-D.H. participated in the discussions and the reviews of the manuscript.

\vspace{.5cm} \noindent {\large \textbf{Additional Information}}

\vspace{.01cm} \noindent \textbf{Competing financial interests:} The authors declare no competing financial interests.


\begin{thebibliography}{99}
%\bibitem{} %\bibitem{} \expandafter\ifx\csname url\endcsname\relax

%\fi
%\expandafter\ifx\csname
%urlprefix\endcsname\relax

%\fi
\providecommand{\bibinfo}[1]{#1} \providecommand{\eprint}[1][]{\url{#1}}

\bibitem{J. Rarity} Rarity, J. G. \emph{et al}. Two-photon interference in a Mach-Zehnder interferometer. \emph{Phys. Rev. Lett.} \textbf{65}, 1348
(1990).

\bibitem{Y. Ji} Ji, Y. \emph{et al}. An electronic Mach-Zehnder interferometer. \emph{Nature} (London) \textbf{422}, 415 (2003).

\bibitem{W. Oliver} Oliver, W. D. \emph{et al}. Mach-Zehnder interferometry in a strongly driven superconducting qubit. \emph{Science} \textbf{310}, 1653 (2005).

\bibitem{V. Giovannetti} Giovannetti, V., Lloyd, S. \& Maccone, L. Quantum-enhanced measurements: beating the standard quantum limit. \emph{Science} \textbf{306}, 1330 (2004).

\bibitem{V. Giovannetti1} Giovannetti, V., Lloyd, S. \& Maccone, L. Quantum metrology. \emph{Phys. Rev. Lett.} \textbf{96}, 010401 (2006).

\bibitem{Yixiao Huang} Huang, Y., Zhong, W., Sun, Z. \& Wang, X. Fisher-information manifestation of dynamical stability and transition to
self-trapping for Bose-Einstein condensates. \emph{Phys. Rev. A} \textbf{86}, 012320 (2012).

\bibitem{J. Est} Est\`{e}ve, J.,  Gross, C., Weller, A., Giovanazzi, S. \& Oberthaler, M. K. Squeezing and entanglement in a Bose-Einstein condensate. \emph{Nature} (London) \textbf{455}, 1216 (2008).

\bibitem{L. Pezz} Pezz\`{e}, L. \& Smerzi, A. Entanglement, nonlinear dynamics, and the Heisenberg limit. \emph{Phys. Rev. Lett.} \textbf{102},
100401 (2009).

\bibitem{Brukner} Brukner, \v{C}., Vedral, V. \& Zeilinger, A. Crucial role of quantum entanglement in bulk properties of solids. \emph{Phys. Rev. A}
\textbf{73}, 012110 (2006).

\bibitem{C. Invernizzi} Invernizzi, C., Korbman, M., Venuti, L. C. \& Paris, M. G. A. Optimal quantum estimation in spin systems at criticality. \emph{%
Phys. Rev. A} \textbf{78}, 042106 (2008).

\bibitem{J. Ma1} Ma, J., Huang, Y. X., Wang, X. \& Sun, C. P. Quantum Fisher information of the Greenberger-Horne-Zeilinger state in decoherence
channels. \emph{Phys. Rev. A} \textbf{84}, 022302 (2011).

\bibitem{J. J. Bollinger} Bollinger, J. J., Itano, W. M., Wineland, D. J. \& Heinzen, D. J. Optimal frequency measurements with maximally correlated
states. \emph{Phys. Rev. A} \textbf{54}, R4649 (1996).

\bibitem{J. P. Dowling} Dowling, J. P. Correlated input-port, matter-wave interferometer: Quantum-noise limits to the atom-laser gyroscope. \emph{%
Phys. Rev. A} \textbf{57}, 4736 (1998).

\bibitem{C. A. Sackett} Sackett, C. A. \emph{et al}.. Experimental entanglement of four particles. \emph{Nature} (London) \textbf{404}, 256 (2000).

\bibitem{R. Huesmann} Huesmann, R., Balzer, Ch., Courteille, Ph., Neuhauser, W. \& Toschek, P. E. Single-Atom Interferometry. \emph{Phys. Rev. Lett.}
\textbf{82}, 1611 (1999).

\bibitem{D. Leibfried} Leibfried, D.\emph{et al}. Trapped-ion quantum Simulator: experimental application to nonlinear interferometers. \emph{Phys. Rev. Lett.} \textbf{89}%
, 247901 (2002).

\bibitem{M. R. Andrews} Andrews, M. R., Townsend, C. G., Miesner, H., Durfee, D. S., Kurn, D. M. \& Ketterle, W. Observation of interference
between two Bose condensates. \emph{Science} \textbf{275}, 637-641 (1997).

\bibitem{E. W. Hagley} Hagley, E. W. \emph{et al}. Measurement of the coherence of a Bose-Einstein condensate. \emph{Phys. Rev. Lett.} \textbf{83}, 3112 (1999).

\bibitem{I. I. Bloch} Bloch, I. I., Hansch, T. W, \& Esslinger, T. Measurement of the spatial coherence of a trapped Bose gas at the phase
transition. \emph{Nature} \textbf{403}, 166 (2000).

\bibitem{A. Perrin} Perrin, A. \emph{et al}. Hanbury Brown and Twiss correlations across the Bose-Einstein condensation threshold. \emph{Nature
Phys.} \textbf{8}, 195 (2012).

\bibitem{Yixiao Huang5} Huang, Y., Xiong, H.-N., Sun, Z., and Wang, X. Generation and storage of spin-nematic squeezing in a spinor Bose-Einstein condensate. \emph{Phys. Rev. A} \textbf{92}, 023622 (2015).

\bibitem{Y. Shin} Shin, Y., \emph{et al}. Atom interferometry with Bose-Einstein condensates in a doublewell potential. \emph{Phys. Rev. Lett.} \textbf{92}, 050405 (2004).

\bibitem{T. Schumm} Schumm, T. \emph{et al}. Matter-wave interferometry in a double well on an atom chip. \emph{Nature Phys.} \textbf{1}, 57 (2005).

\bibitem{G. B. Jo} Jo, G.-B. \emph{et al}. Long phase coherence time and number squeezing of two Bose-Einstein condensates on an atom chip. \emph{Phys. Rev. Lett.} \textbf{98}, 030407 (2007).

\bibitem{G. B. Jo1} Jo, G.-B. \emph{et al}. Phase-sensitive recombination of two Bose-Einstein condensates on an atom chip. \emph{Phys. Rev. Lett.} \textbf{98}, 180401 (2007).

\bibitem{Baumga} Baumga\"{r}tner, F. \emph{et al}. Measuring energy differences by BEC interferometry on a chip. \emph{Phys. Rev. Lett.}
\textbf{105}, 243003 (2010).

\bibitem{Pezze} Pezz\'{e}, L., Collins, L., Smerzi, A., Berman, G. \& Bishop, A. Sub-shot-noise phase sensitivity with a Bose-Einstein condensate
Mach-Zehnder interferometer. \emph{Phys. Rev. A} \textbf{72}, 043612 (2005).

\bibitem{J. Grond} Grond, J., Hohenester, U., Mazets, I. \& Schmiedmayer, J. Atom interferometry with trapped Bose-Einstein condensates: impact of
atom-atom interactions. \emph{New J. Phys.} \textbf{12}, 065036 (2010).

\bibitem{J. Grond1} Grond, J., Hohenester, U., Schmiedmayer, J. \& Smerzi, A. Mach-Zehnder interferometry with interacting trapped Bose-Einstein
condensates. \emph{Phys. Rev. A} \textbf{84}, 023619 (2011).

\bibitem{C. Lee} Lee, C. Adiabatic Mach-Zehnder interferometry on a quantized Bose-Josephson junction. \emph{Phys. Rev. Lett.} \textbf{97},
150402 (2006).

\bibitem{Berrada T} Berrada T. \emph{et al}. Integrated Mach--Zehnder interferometer for Bose-Einstein condensates. \emph{Nature Commun.} \textbf{4}, 2077 (2013).

\bibitem{Yixiao Huang1} Huang, Y., Sun, Z. \& Wang, X. Atom-number fluctuation and macroscopic quantum entanglement in dipole spinor condensates. \emph{Phys. Rev. A} \textbf{89}, 043601 (2014).

\bibitem{S. Yi} Yi, S. \& Pu, H. Magnetization, squeezing, and entanglement in dipolar spin-1 condensates. \emph{Phys. Rev. A} \textbf{73}, 023602
(2006).

\bibitem{Yixiao Huang3} Huang, Y., Zhang, Y., L\"{u}, R., Wang, X. \& Yi, S. Macroscopic quantum coherence in spinor condensates confined in an
anisotropic potential. \emph{Phys. Rev. A} \textbf{86}, 043625 (2012).

\bibitem{S. Yi2} Yi, S., You, L. \& Pu, H. Quantum phases of dipolar spinor condensates. \emph{Phys. Rev. Lett.} \textbf{93}, 040403 (2004).

\bibitem{C. K. Law} Law, C. K., Pu, H. \& Bigelow, N. P. Quantum spins mixing in spinor Bose-Einstein condensates. \emph{Phys. Rev. Lett.} \textbf{81}, 5257 (1998).

\bibitem{S. Yi3} Yi, S., Mustecaplio\v{g}lu, O. E., Sun, C. P. \& You, L. Single-mode approximation in a spinor-1 atomic condensate. \emph{Phys. Rev. A%
} \textbf{66}, 011601 (2002).

%\bibitem{Stamper-Kurn} Stamper-Kurn, D. M., Andrews, M. R., Chikkatur, A.
%P., Inouye, S., Miesner, H.-J., Stenger, J. \& Ketterle, W. Optical
%Confinement of a Bose-Einstein Condensate. \emph{Phys. Rev. Lett.} \textbf{80%
%}, 2027 (1998).

%\bibitem{C. W. Helstrom} Helstrom, C. W. Quantum Detection and Estimation
%Theory (Academic Press, New York, 1976).
%
%\bibitem{A. S. Holevo} Holevo, A. S. Probabilistic and Statistical Aspects
%of Quantum Theory (North-Holland, Amsterdam, 1982).
%
%\bibitem{M. Hayashi} Hayashi, M. Quantum Information: An Introduction
%(Springer- Verlag, Berlin, 2006).

\bibitem{M. Kitagawa} Kitagawa, M. \& Ueda, M. Squeezed spin states. \emph{Phys. Rev. A}, \textbf{47}, 5138 (1993).

\bibitem{J. Ma} Ma, J., Wang, X., Sun, C. P. \& Nori, F. Quantum spin squeezing. \emph{Phys. Rep.}, \textbf{509}, 89 (2011).

\bibitem{X. Wang} Wang, X. \& Sanders, B. C. Spin squeezing and pairwise entanglement for symmetric multiqubit states. \emph{Phys. Rev. A}, \textbf{%
68}, 012101 (2003).

\bibitem{T. J. Osborne} Osborne, T. J. \& Nielsen, M. A. Entanglement in a simple quantum phase transition. \emph{Phys. Rev. A}, \textbf{66}, 032110
(2002).

\bibitem{J. Vidal} Vidal, J., Palacios, G. \& Mosseri, R. Entanglement in a second-order quantum phase transition. \emph{Phys. Rev. A}, \textbf{69},
022107 (2004).
\end{thebibliography}
\end{document}